\documentclass[epj,final]{svjour}
\usepackage{latexsym,amsmath}
\usepackage{graphics}
\usepackage{graphicx}
\usepackage{amssymb}

\begin{document}

\title{Microwave Properties of Ba$_{0.6}$K$_{0.4}$BiO$_3$ Crystals}

\author{S. Fricano\inst{1} \and M. Bonura\inst{1} \and A. Agliolo~Gallitto\inst{1}
\and{M. Li~Vigni\inst{1} \and L. A. Klinkova\inst{2} \and N. V.
Barkovskii\inst{2}} \institute{INFM and Dipartimento di Scienze
Fisiche e Astronomiche, Universit$\mathrm{\grave{a}}$ di Palermo,
Via Archirafi 36, I-90123 Palermo (Italy) \and Institute of Solid
State Physics, Chernogolovka, Moscow District 142432, Russia}
}%

\abstract{We report on field-induced variations of the microwave
surface resistance at 9.6 GHz of Ba$_{0.6}$K$_{0.4}$BiO$_3$
crystals. Energy losses have been investigated as a function of
the static magnetic field in the range of temperatures 4.2~K
$\div~ T_c$. By analyzing the experimental results in the
framework of the Coffey and Clem model we determine the
temperature dependence of the first-penetration field, upper
critical field and depinning frequency. The results show that the
pinning energy of this bismuthate superconductor is weaker than
those of cuprates.
 \PACS{
      {74.25.Ha}{Magnetic properties}\and
      {74.25.Nf}{Response to electromagnetic fields (nuclear magnetic
      resonance, surface impedance, etc.)}\and
      {74.25.Op}{Mixed state, critical fields, and surface sheaths}
     }
}

\maketitle

\section{Introduction}
The bismuthate superconductor $\mathrm{Ba_{0.6}K_{0.4}BiO_3}$
(BKBO) is often classified as a high-temperature superconductor
because of the relatively high critical temperature, $T_c \simeq
30$~K well above that of conventional superconductors. The
interest in the study of the BKBO compound is due to the fact
that, although it does not contain copper, some characteristics
are similar to those of cuprate superconductors: it is an oxide
with perovskite structure, low charge-carrier density and
superconductivity near a metal-insulator
transition~\cite{barilo,lee,hinks}. Despite such similarities,
some marked differences can be highlighted: bismuthates are
nonmagnetic, and they have three-dimensional structures rather
than layered two-dimensional ones characteristic of
cuprates~\cite{klink}. Though several properties of BKBO have been
explained in the framework of the BCS theory~\cite{sato,hinks2},
with a single type of carrier pairing, anomalies such as, e.g.,
the upward curvature at temperatures near $T_c$ of the upper
critical field, cannot be explained in the weak-coupling limit
\cite{affronte,samuely,gant,hall,miura}. A complication in
interpreting the experimental results is related to the intrinsic
difficulty of producing BKBO crystals with a highly uniform K-ion
content. Indeed, the presence of micro-domains with different
potassium concentration, which could strongly affect several
physical properties, seems to be an intrinsic feature of the
available BKBO samples.

The magnetic properties of BKBO have been investigated by
different techniques such as magnetic susceptibility, specific
heat and thermal conductivity
measurements~\cite{barilo,gant,yang}. The results have highlighted
very small values of the pinning energy, consistent with those
measured in other bismuthate superconductors~\cite{maley}. This
small pinning energy is responsible for the enhanced dissipation
and the low critical current observed in the bismuthate
superconductors. Usually, the pinning characteristics are
extracted from the field dependence of the critical current;
however, since the critical current of high-$T_c$ superconductors
at low temperatures is too large to be measured by direct-contact
method, the study of the electromagnetic ($em$) response of
superconductors in the mixed state is a more convenient method.
Measurements of high-frequency $em$ response are particularly
suitable to investigate vortex dynamics because they probe the
vortex response at very low currents, when the vortices undergo
reversible oscillations~\cite{golo}. The most commonly method to
study the high-frequency $em$ response consists in measuring the
microwave ($mw$) surface impedance, $Z_s = R_s -
iX_s$~\cite{golo,mishagrande}. The real component, $R_s$, is
proportional to the energy losses while the imaginary component,
$X_s$, is related to the field penetration depth.  The different
vortex states, in the different regions of the $H$-$T$~phase
diagram, determine the temperature and field dependencies of
$Z_s$; so, measurements of $Z_s(H, T)$ may provide important
information on the fluxon dynamics in the different regimes of
motion~\cite{golo,owl92,cc,bra}.

In this paper, we report experimental results of field-induced
variations of $R_s$ in a single crystal of BKBO. The experimental
results allow measuring the values of the magnetic field at which
fluxons start to penetrate the sample and the temperature
dependence of the upper critical field. The experimental curves of
$R_s(H, T)$ are well accounted for in the framework of the Coffey
and Clem model~\cite{cc} in the whole range of temperatures
investigated ($4.2~\mathrm{K} \div T_c$). By fitting the
experimental data, using this model, we have deduced the
temperature dependence of the depinning frequency. Near $T_c$, the
$mw$ current induces fluxons moving in the flux-flow regime and
the depinning frequency is almost zero; it remains roughly zero
for $\approx 8$~K below $T_c$; on further decreasing the
temperature, the value of the depinning frequency increases up to
$\sim 20$~GHz at $T = 4.2$~K. This value confirms that even at low
temperatures the pinning energy is small; so that, the motion of
weakly-pinned fluxons, induced by the $mw$ current, gives rise to
relatively large energy losses.
\section{Experimental and Sample}
The field-induced variations of $R_s$ have been studied in a
single crystal of BKBO of nearly cubic shape, with about 1~mm
edge; it undergoes a superconducting transition at $T_c \approx
32$~K, of width $\approx 2.5$~K. The sample has been synthesized
at the Institute of Solid State Physics of Russian Academy of
Science. It has been produced by electrolysis of
KOH-Ba(OH)$_{2}$-Bi$_{2}$O$_{3}$ melt, under conditions of
stationary mass transfer~\cite{klinkova}. The
$\mathrm{Ba_{1-x}K_{x}BiO_3}$ crystals produced with this method
have a potassium concentration near to the optimal value, $x
\approx 0.4$. However, they exhibit a wide superconducting
transition and a considerable residual $mw$ surface resistance at
low temperatures. These features can be ascribed to an
inhomogeneous distribution of K ions in the crystals~\cite{misha}.

The $mw$ surface resistance has been measured using the cavity
perturbation technique~\cite{mishagrande}. The copper cavity, of
cylindrical shape with golden-plated walls, is tuned in the
TE$_{011}$ mode resonating at 9.6~GHz ($Q \approx 40,000$ at $T =
4.2$~K). It is placed between the poles of an electromagnet, which
generates static magnetic fields, $H_0$, up to $\approx 10$~kOe.
Two additional coils, independently fed, allow compensating the
residual field and working at low magnetic fields. The sample is
put in the center of the cavity by a sapphire rod, in the region
of maximum $mw$ magnetic field. The sample and field geometry is
shown in Fig.~1~(a); in this geometry the $mw$ current induces a
tilt motion of all the fluxons present in the sample, as shown in
Fig.~1~(b). The $Q$-factor of the cavity is measured by means of
an $hp$-8719D Network Analyzer. The surface resistance of the
sample is proportional to ($1/Q_L - 1/Q_U$), where $Q_U$ is the
$Q$-factor of the empty cavity and $Q_L$ is that of the loaded
cavity. In order to disregard the geometric factor of the sample,
it is convenient to normalize the deduced values of $R_s$ to the
value of the surface resistance at a fixed temperature in the
normal state, $R_n$. In particular, we have normalized all the
$R_s$ values here reported to the value of $R_n$ at $T = 32$~K and
$H_0 = 0$. Measurements have been performed as a function of the
temperature and the static magnetic field.
\begin{figure}[t]
\centering
\includegraphics[width=7cm, height=4cm]{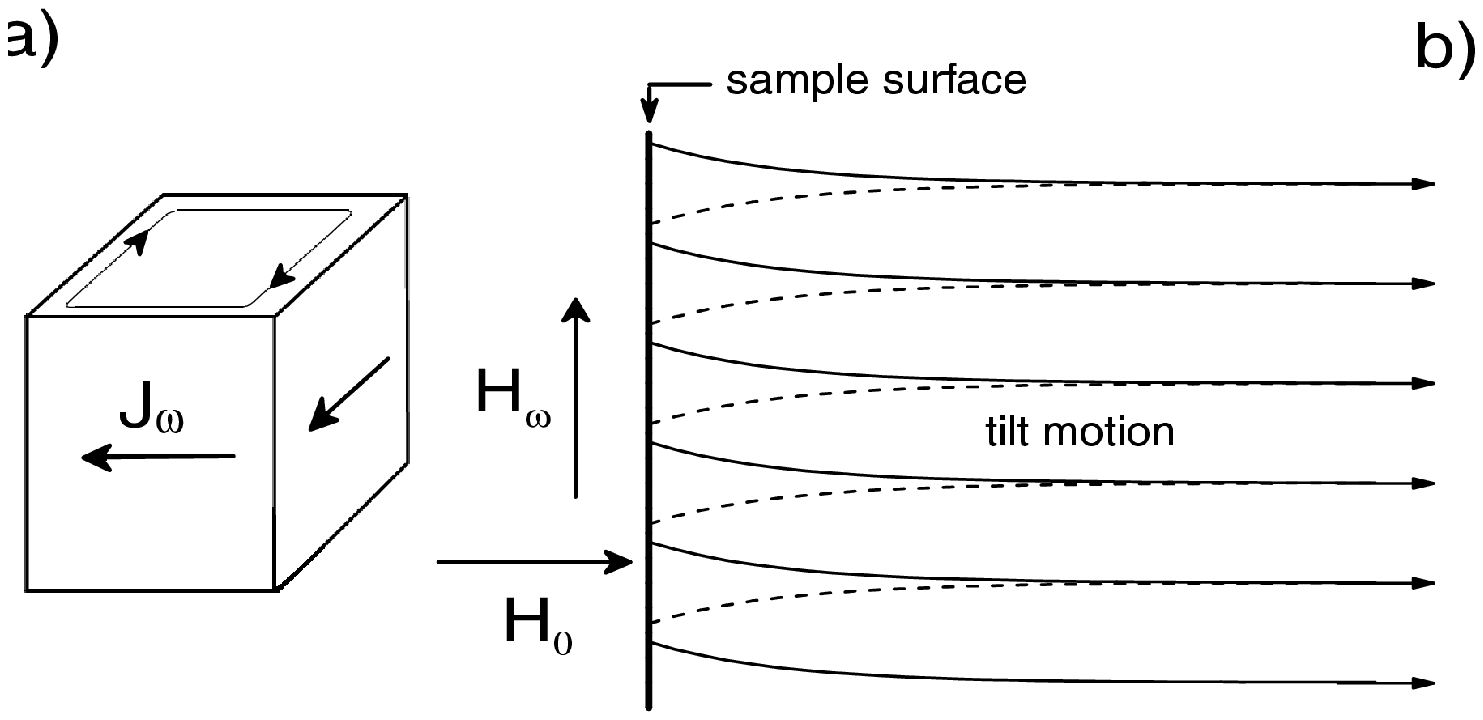}
\caption{a) Sample and field geometry. b) Fluxon motion induced by
the $mw$ current; the arrows indicate the flux lines.}
\end{figure}

Fig.~2 shows the temperature dependence of the normalized surface
resistance, at different values of the static field. The results
have been obtained according to the following procedure: the
sample was zero-field cooled (ZFC) down to 4.2~K, then $H_0$ was
set at a given value, which was kept constant during the time in
which the measurements in the temperature range $4.2 \div 32$~K
have been performed. As expected, on increasing $H_0$, the
$R_s(T)$ curve broadens and shifts toward lower temperatures.
Measuring the temperature and field at which $R_s/R_n = 1$, we
have estimated $dH_{c2}(T)/dT|_{T_c} \sim 5 $ kOe/K, in agreement
with the results reported in the literature \cite{affronte}.
\begin{figure}[b]
\centering
\includegraphics[width=8cm]{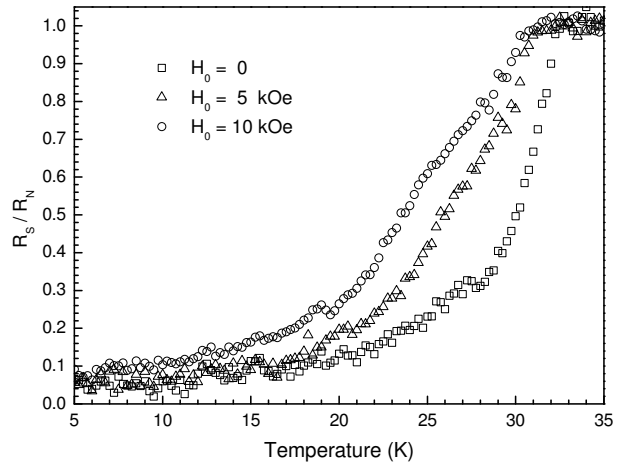}
\caption{Normalized values of the $mw$ surface resistance as a
function of the temperature, for different values of $H_0$.}
\end{figure}

$R_s$ has also been investigated as a function of the magnetic
field, in the range $0\leq H_0\leq 10$~kOe, at fixed values of
temperature. Each measurement has been performed, in the ZFC
sample, sweeping the external field up to 10~kOe and,
successively, decreasing it down to zero, at constant temperature.
Fig.~3 shows $R_s/R_n$ as a function of $H_0$ at $T = 4.2$~K. Open
and solid symbols refer to results obtained on increasing and
decreasing $H_0$, respectively; no magnetic hysteresis has been
observed. In order to better highlight the low-field dependence of
$R_s/R_n$, in the inset we report the data in a semi-log plot. As
one can see, the curve is field independent in the range $0 \div
H_p$, showing that, in this range, the external magnetic field
does not induce variations of the surface resistance. Since the
sample was ZFC, $H_p$ should mark the first-penetration field.
\begin{figure}[tb]
\centering
\includegraphics[width=8cm]{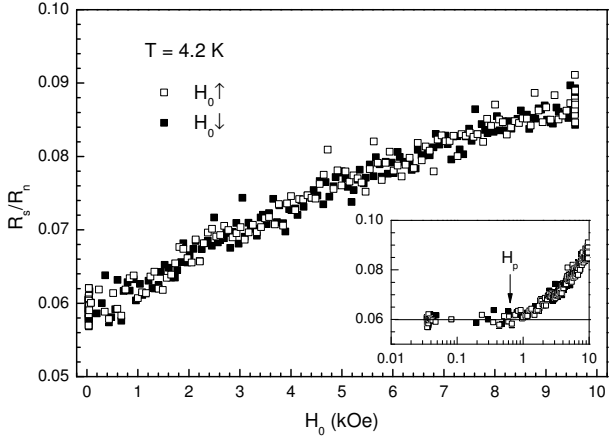}
\caption{Normalized values of $R_s$ as a function of the static
magnetic field at $T = 4.2$~K. Open and solid symbols refer to
results obtained on increasing and decreasing field, respectively.
The arrow in the inset indicates the first-penetration field,
$H_p$.}
\end{figure}

Fig.~4 shows the normalized values of the surface resistance as a
function of the static magnetic field, at different values of the
temperature in the range 10 $\div$ 30 K; symbols show the
experimental data, lines are the best-fit curves of data obtained
with the procedure described in the next section. As one can see,
on increasing the temperature the field-induced variations of
$R_s$ increase.
\begin{figure}[t]
\centering
\includegraphics[width=8cm]{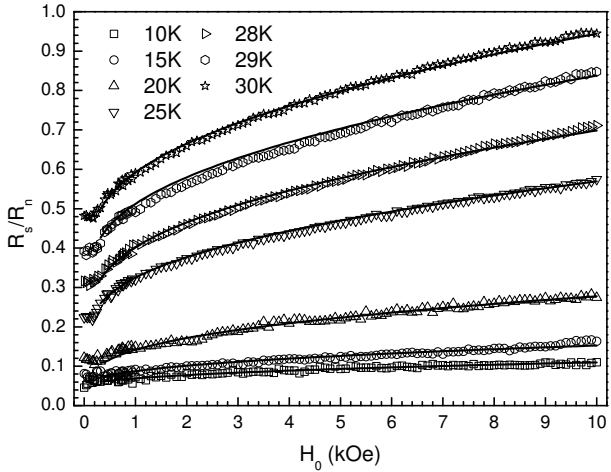}
\caption{Normalized values of $R_s$ as a function of the static
magnetic field, for different values of temperature. The lines are
the best-fit curves, obtained as described in Section 3.}
\end{figure}
By analyzing the experimental data of Fig.s~3 and 4 obtained at
low applied fields, we have deduced the temperature dependence of
the first-penetration field, which is shown in Fig.~5. As one can
see, on decreasing the temperature $H_p$ increases monotonically
even at low temperatures; this behavior has already been reported
by Hall $et~al.$ \cite{hall}. The value of $H_p(0)$ results
slightly larger than the $H_{c1}(0)$ values reported in the
literature for BKBO crystals \cite{barilo,hall}, suggesting that
weak surface-barrier effects are present in our sample.
\begin{figure}[h]
\centering
\includegraphics[width=8cm]{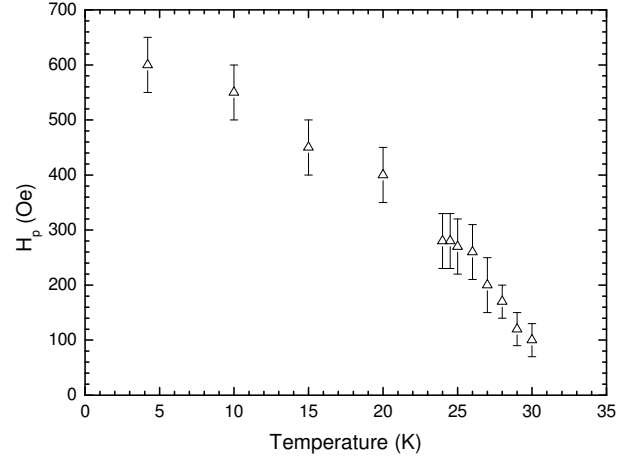}
\caption{Temperature dependence of the first-penetration field,
$H_p$. The symbols show the $H_p$ values deduced from the
$R_s(H,T)$ curves, as the $H_0$ values at which $R_s$ starts
increasing.}
\end{figure}

In Fig.~6 we report the normalized values of the surface
resistance as a function of the static magnetic field at
temperatures near $T_c$. The symbols show the experimental data,
the lines have been obtained by fitting the data with the
procedure described in the next section. At temperatures close to
$T_c$, the sample goes into the normal state at $H_0$ values that
our experimental apparatus can easily supply. At a fixed
temperature, measuring the value of $H_0$ at which $R_s$ reaches
the normal-state value we have determined $H_{c2}(T)$; the deduced
values are consistent with those obtained from the $R_s$-$vs$-$T$
curves and those reported in the literature for BKBO crystals
\cite{affronte}.
\begin{figure}[tb]
\centering
\includegraphics[width=8cm]{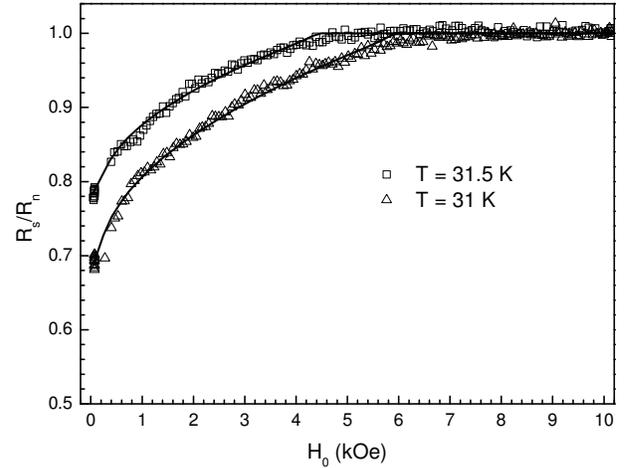}
\caption{Normalized values of $R_s$ as a function of $H_0$, for
two values of temperature close to $T_c$. The lines are the
best-fit curves, obtained as described in Section 3.}
\end{figure}
\section{Discussion}
Microwave losses induced by the static magnetic field in
superconductors in the mixed state are mainly due to the motion of
fluxons and, at temperatures near $T_c$, to the very presence of
vortices, which bring along normal fluid in their core
\cite{cc,agl97}. For ZFC samples, on increasing the static
magnetic field, the surface resistance begins to increase when
$H_0$ reaches the value at which fluxons start penetrating the
sample. So, in our results the value of the field indicated with
$H_p$ (see inset in Fig. 3) marks the first-penetration field.
From Fig. 5, one can see that the values of $H_p$ obtained at the
lowest temperature investigated are slightly larger than the
values of $H_{c1}(0)$ reported in the literature for BKBO
crystals~\cite{barilo}, suggesting that in our sample weak
surface-barrier effects are present. However, since at
temperatures close to $T_c$ the surface-barrier effects are
negligible, in this range of temperatures, $H_p$ should coincide
with $H_{c1}$. As one can see from Fig.~5, the $H_p(T)$ curve does
not exhibit upward curvature near $T_c$, in contrast to what
occurs in the $H_{c2}(T)$ curve of BKBO
\cite{affronte,samuely,gant,miura}. This is a peculiar property of
BKBO; indeed, though in conventional superconductors $H_{c1}$ and
$H_{c2}$ near $T_c$ have the same temperature dependence, in BKBO
they show different behavior. Studies reported in the literature
on the origin of this property have suggested that the upward
curvature of $H_{c2}(T)$ is due to intrinsic disorder in the
crystal structure of BKBO~\cite{gant}. The reason for which this
does not influence $H_{c1}(T)$, leading to the absence of the
upward curvature, is still unclear.

The field-induced variations of $R_s$ in the mixed state have been
studied by different authors~\cite{golo,cc,bra}. Coffey and Clem
(CC) have developed a comprehensive theory for the $em$ response
of type-II superconductors in the mixed state, in the framework of
the two-fluid model~\cite{cc}. The theory applies for $H_0 >
2H_{c1}$, when the induction field inside the sample can be
supposed as generated by an uniform density of fluxons; in this
case $H_0 \approx B_0 = n \phi_0$, where $\phi_0$ is the flux
quantum and $n$ is the vortex density. We will show that our
results can be well accounted for by the CC model taking into
consideration that in different temperature ranges the $mw$
current induces fluxons moving in different regimes.

In the London local limit, the surface impedance is proportional
to the complex penetration depth of the $em$ field,
$\widetilde{\lambda}$. In particular
\begin{equation}\label{rs}
R_s = -\frac{4 \pi \omega}{c^2} \Im(\widetilde{\lambda}).
\end{equation}
In the CC model $\widetilde{\lambda}$ is calculated by taking into
account the effects of the fluxon motion and the very presence of
vortices. In the linear approximation, $H(\omega)\ll H_0$, the
following expression for $\widetilde{\lambda}$ has been obtained
\cite{cc}:
\begin{equation}\label{lamtil}
\widetilde{\lambda}^2=\frac{\lambda^2+\widetilde{\delta}_{v}^2}
{1-2i\lambda^2/\delta^2},
\end{equation}
where $\widetilde{\delta}_v$ is the effective complex skin depth
arising from vortex motion; $\lambda$ and $\delta$, the London
penetration depth and the normal-fluid skin depth, are given by
\begin{equation}\label{lamda0}
\lambda = \frac{\lambda_0}{\sqrt{(1-w)(1- B_0 /H_{c2})}},
\end{equation}
\begin{equation}\label{delta0}
\delta = \frac{\delta_0}{\sqrt{1-(1-w)(1- B_0 /H_{c2})}},
\end{equation}
here $\lambda_0$ is the London penetration depth at  $T = 0$;
$\delta_0$ is the normal-fluid skin depth at $T = T_c$; $w$ is the
fraction of normal electrons at $H_0 = 0$, in the Gorter and
Casimir two-fluid model $w = (T/T_c)^4$. \\The penetration depth
$\widetilde{\delta}_v$ can be written in terms of the two
characteristic lengths, $\delta_f$ and $\lambda_c$, arising from
the contributions of the viscous and the restoring-pinning forces,
respectively:
\begin{equation}\label{deltav}
\frac{1}{\widetilde{\delta}_{v}^2}=\frac{1}{\lambda_{c}^2}-\frac{2i}{\delta_{f}^2},
\end{equation}
where
\begin{equation}\label{labdac}
\lambda_{c}^2 = \frac{B_0 \phi_0}{4 \pi k_p},
\end{equation}
\begin{equation}\label{deltaf}
\delta_{f}^2 = \frac{B_0 \phi_0}{2 \pi \omega \eta},
\end{equation}
with $k_p$ the restoring-force coefficient and $\eta$ the
viscous-drag coefficient. In s-wave superconductors, such as BKBO
\cite{Yokoya,bansil}, the viscous-drag coefficient is given by the
Bardeen-Stephen expression \cite{bs}
\begin{equation}\label{eta}
\eta= \frac{\phi_0 H_{c2}}{\rho_n c^2},
\end{equation}
where $\rho_n$ is the normal-state resistivity.

The effectiveness of the two terms in Eq.~(\ref{deltav}) depends
on the ratio $\omega_c = k_p/\eta$, which defines the depinning
frequency. In terms of $\omega_c$, Eq.~(\ref{deltav}) can be
written as
\begin{equation}\label{deltav2}
\frac{1}{\widetilde{\delta}_{v}^2}=\frac{1}{4
\delta_{f}^2}\left(\frac{\omega_c}{\omega}-\frac{i}{2}\right).
\end{equation}
When the frequency of the $em$ wave, $\omega$, is much larger than
$\omega_c$ the contribution of the viscous-drag force predominates
and the induced $em$ current makes fluxons move in the flux-flow
regime. On the contrary, for $\omega \ll \omega_c$ the motion of
fluxons is ruled by the restoring-pinning force.

It is worth noting that the above-mentioned expressions have been
obtained by supposing that the $em$ current induces compressional
waves of the fluxon-line lattice within the field penetration
depth; as a consequence, $B_0$ is the induction field within
$\widetilde{\lambda}$. In the field geometry of our experimental
apparatus, the $mw$ current induces a $tilt$ motion of all the
fluxons present in the sample. On the other hand, Brandt
\cite{bra} has shown that the $compressional$ and $tilt$ motion
can be described by the same formalism, and the same complex
penetration depth results since the moduli for long-wavelength
compression and tilt are approximately equal. So, we can consider
valid the expressions obtained in the CC model provided that $B_0$
be the induction field averaged over the whole sample.

Eqs. (1)-(9) allow calculating the field-induced variations of the
surface resistance, at fixed values of the temperature. However,
they cannot account for the residual surface resistance, measured
at zero dc field, which is mainly related to the sample
inhomogeneity. In order to compare the expected and experimental
results, at fixed temperature values, we have calculated the
field-induced variations of $R_s/R_n$ normalized to the maximal
value:

\begin{equation}\label{delta Rs}
    \frac{[R_s(H, T)-R_s(0, T)]/R_n}{1-R_s(0,T)/R_n}=\frac{2\Im[\widetilde{\lambda}(H, T)-
    \widetilde{\lambda}(0, T)]/\delta_0}{1-2\Im(\widetilde{\lambda}(0, T))/\delta_0},
\end{equation}
where the left-side term can be determined from the experimental
data and the right-side term is the expected one, $\delta_0/2$
corresponds to the value of $\Im(\widetilde{\lambda})$ in the
normal
state.\\
By considering the values of the characteristic parameters of BKBO
crystals, it can be shown that, except at temperatures very close
to $T_c$, $\Im[\widetilde{\lambda}(0, T)]\ll\delta_0/2$.
Therefore, from Eq.~(\ref{delta Rs}) it results:\\
\begin{eqnarray}
    \lefteqn{\frac{R_s(H, T)}{R_n} \simeq \frac {R_s(0, T)}{R_n}+ {} }\nonumber\\
    & & {}+\left[1-\frac {R_s(0,T)}{R_n}\right]
    \frac{2\Im[\widetilde{\lambda}(H, T)-\widetilde{\lambda}(0, T)]}{\delta_0}.
\label{Rsfit}
\end{eqnarray}
\\
Eq.~(\ref{Rsfit}) has been used to fit the experimental
data\footnote{The approximation used to get Eq. (\ref{Rsfit}) has
been verified by numerical calculations using the values of the
characteristic parameters of BKBO ($H_{c1}, H_{c2}, T_c,
\lambda_0$ and $\delta_0$); it gives rise to corrections much
smaller than the experimental uncertainty up to a few tenths of K
below $T_c$.}. It is worth remarking that only the experimental
results obtained for $H_0 > H_p$ have been fitted.

At temperatures near $T_c$ the induced $mw$ current makes fluxons
move in the flux-flow regime, so we can assume $\omega \gg
\omega_c \approx 0$. In this case, $\delta_v^2 \approx i
\delta_f^2/2$ and, in order to perform a comparison between the
experimental and the expected $R_s(H, T)$ curves, besides $R_s(0,
T)/R_n$, only two parameters are necessary: the ratio $\lambda_0/
\delta_0$ and $H_{c2}(T)$. Since, at temperatures near $T_c$, the
$H_{c2}(T)$ values can be directly deduced from the $R_s(H,T)$
curves, the only free parameter necessary for fitting the
experimental data, obtained near $T_c$, is $\lambda_0/\delta_0$.
The lines in Fig.~6 have been obtained by Eq.~(11) using for
$H_{c2}(T)$ and $R_s(0, T)/R_n$ the values deduced from the
experimental results, letting them vary within the experimental
uncertainty, and taking $\lambda_0/\delta_0$ as free parameter.
The best-fit curves of Fig.~6 have been obtained with
$\lambda_0/\delta_0$ = 0.04; however, we have found that the
expected results are little sensitive to variations of
$\lambda_0/\delta_0$ as long as it takes on values of the order of
$10^{-2}$. It is easy to see that $\lambda_0/\delta_0 =
\sqrt{\omega\tau/2}$, where $\tau$ is the scattering time of the
normal electrons; so the corresponding value of $\tau$ is of the
order of $10^{-14}$~s. This value agrees with the ones reported in
the literature for BKBO and is smaller than that reported for
cuprate superconductors \cite{mishagrande,misha,waldram}.

For $T\leq$ 30 K, the $H_{c2}(T)$ values cannot be directly
deduced from the experimental $R_s(H, T)$ results. So, by keeping
for $\lambda_0/\delta_0$ the same value obtained at temperatures
near $T_c$, the parameters necessary to fit the experimental data
are $\omega_c$ and $H_{c2}(T)$. Since $H_{c2}(T)$ of our sample at
temperatures near $T_c$ agrees quite well with the values reported
in the literature for BKBO crystals \cite{affronte}, we have
assumed that even at lower temperatures the values of $H_{c2}(T)$
of our sample are consistent with those reported in the
literature. The lines reported in Fig.~4 have been obtained by
Eq.~(\ref{Rsfit}) taking $\lambda_0/\delta_0 = 0.04$, $\omega_c$
as free parameter and letting $H_{c2}(T)$ vary compatibly with the
values reported in the literature for BKBO crystals. As one can
see, the expected results describe quite well the experimental
data in the whole range of temperatures investigated. In Fig.~7 we
report the temperature dependence of the upper critical field as
deduced by the fitting procedure.
\begin{figure}[tb]
\centering
\includegraphics[width=8cm]{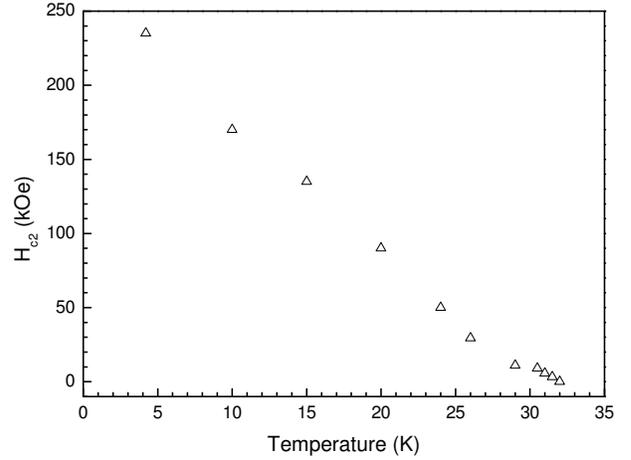}
\caption{Values of the upper critical field determined as best-fit
parameters of the experimental data, using the procedure described
in the text.}
\end{figure}
\begin{figure}[tb]
\centering
\includegraphics[width=8cm]{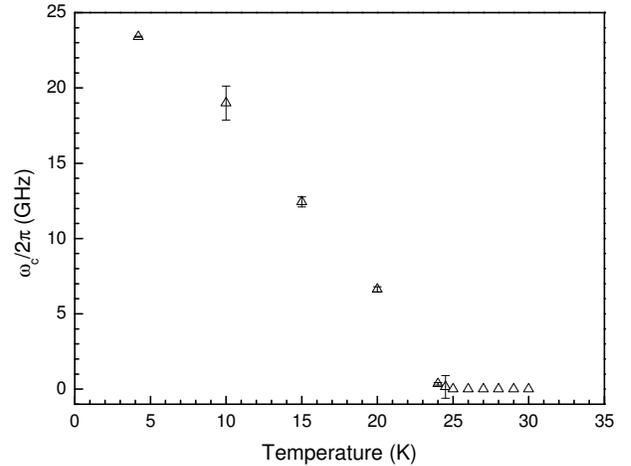}
\caption{Depinning frequency as a function of the temperature,
determined as described in the text.}
\end{figure}

The values of $\omega_c(T)$ which best fit our experimental data
are plotted in Fig.~8. In the range of temperatures 25~K~$ < T <
T_c$, $\omega_c$ is zero, within the experimental accuracy; it
means that for $T > 25$~K the condition $\omega\gg \omega_c$ is
verified and the fluxons, under the action of the $mw$ current,
move in the flux-flow regime. On decreasing the temperature the
depinning frequency increases, reaching the value
$\omega_c/2\pi\simeq 23.4$ GHz at $T = 4.2$~K; this value is lower
than the ones reported in the literature for cuprate
superconductors. From this value of $\omega_c$, using
Eq.~(\ref{eta}) we obtain $k_p \simeq 3.5 \cdot 10^3$ N/m$^2$ at
$T = 4.2$~K. As a consequence of the small value of $\omega_c$, we
obtained a pinning constant smaller than that of cuprate
superconductors, in agreement with what reported in the literature
for BKBO \cite{barilo}. As it is well known, the pinning constant
is determined by the interaction between vortices and pinning
centers as well as by the vortex elasticity. Whereas the effects
of the pinning centers may depend on the investigated sample, the
vortex elasticity is an intrinsic property of the compound and,
usually, its contribution is weaker than the previous one. The
value we found for $k_p$ confirms that, even at low temperatures,
the pinning energy in BKBO is smaller than in cuprate
superconductors, suggesting that the density of pinning centers
and the vortex elasticity are too small to prevent $mw$
absorption.

\section{Conclusions}
In this paper we have reported a detailed study of the
field-induced variations of the $mw$ surface resistance in a
crystal of BKBO. The $mw$ energy losses have been investigated as
a function of the temperature and the static magnetic field. We
have shown that the experimental results can be well accounted for
in the framework of the Coffey and Clem model provided that
different regimes of fluxon motion are hypothesized in different
ranges of temperature. We have determined the temperature
dependence of the first-penetration field and the upper critical
field. The values of $H_p(T)$ have been directly deduced from the
experimental data in the whole range of temperatures investigated.
The temperature dependence of the upper critical field has been
obtained determining $H_{c2}(T)$ as fitting parameter.
Furthermore, by fitting the experimental data we have determined
the temperature dependence of the depinning frequency in the whole
range of temperatures investigated. For $\sim 8$ K below $T_c$, we
have found $\omega_c/\omega\ll 1$, that means that the $mw$
current induces fluxons moving in the flux-flow regime. On further
reducing the temperature, the depinning frequency increases,
reaching the value $\omega_c/2\pi\simeq$ 23.4 GHz at $T = 4.2$ K.
This finding confirms that in this superconductor the pinning
effects are weaker than in the cuprate superconductors and
accounts for the enhanced energy losses and the low value of the
critical current, reported in the literature for bismuthate
superconductors.

\section*{Acknowledgements} The authors are very glad to thank E. H. Brandt, G. E.
Tsydynzhapov, I. Ciccarello and M. Guccione for critical reeding
of the manuscript and helpful suggestions; G. Lapis and G. Napoli
for technical assistance. Work partially supported by the
University of Palermo (grant Coll. Int. Li Vigni).

\begin{flushright}\today \end{flushright}

\end{document}